\documentclass[a4paper,12pt,english,german]{article}

\usepackage{mathtools}
\usepackage{graphicx}

\begin{document}

\centerline{\bf  A top-down versus a bottom-up hidden-variables
description}

\centerline{\bf of the Stern-Gerlach experiment}

\bigskip

\centerline{M. Arsenijevi\' c$^{\dag}$, J. Jekni\' c-Dugi\'
c$^{\ast}$, M. Dugi\' c$^{\dag}$}

\smallskip

$^{\ast}${Department of Physics, Faculty of Science, Ni\v s,
Serbia}

$^{\dag}${Department of Physics, Faculty of Science, Kragujevac,
Serbia}

\bigskip

{\bf Abstract} We employ the Stern-Gerlach experiment to highlight the basics of a minimalist,
non-interpretational top-down approach to quantum foundations. Certain benefits of the
here highlighted "quantum structural studies" are detected and discussed.  While the top-down approach
can be described without making any reference to the fundamental structure of a closed system,
the hidden variables theory \' a la Bohm proves to be more subtle than it is typically regarded.

\bigskip

{\bf 1. Introduction}

\bigskip

Reductionist, i.e. bottom up, thinking is prominent in physics:
Features and dynamics of the structural components (subsystems),
all the way down to the elementary particles, are assumed to
exhaust the description of the features and dynamics of the whole.
Huge success of application of physical theories seem to weaken
the possible merits of the question: What might be wrong with the
physical reductionism?

However, there are indications that this "wrong" is rather subtle
considering that reductionism may  be not the "whole story". On
one hand, efficient physical description of many-particle systems
is lacking [1]. On the other hand, even the orthodox "quantum"
systems may hide subtleties in their structural description, given
entanglement and other features of "quantum wholeness". It may be
more realistic to allow native variables for quantum composites.

In the context of the universally valid and complete quantum
mechanics not extended by any additional rules or (e.g. interpretational) assumptions [2]:

\noindent "{\it Without further physical assumption, no partition
has an ontologically superior status with respect to any other.}"

\noindent as well as [3]:

\noindent "{\it However, for many macroscopic systems, and in
particular for the universe as a whole, there may be no natural
split into distinguished subsystems and the rest, and another way
of identifying the naturally decoherent variables is required.}".

\noindent The apparent lack of  preferred structure (partition
into subsystems) of a closed quantum system yields a top-down
approach to quantum structures, cf. e.g. [4] (and references
therein).

In the context of the universally valid but incomplete quantum
theory, the things may look the opposite. For example, in the de
Broglie-Bohm quantum theory, the particles are assumed to define
the physically fundamental (ontic) structure of the Universe [5].
This kind of structuralism is bottom-up that {\it supplements} the
standard quantum mechanical formalism.

Hence the foundational character of the quantum structure studies
and particularly of the topical questions highlighting this Volume
that can be shortly expressed as "How components relate to a
composite?". Scientific relevance of this question stands even
without any reference to "applications", since majority of the
working physicists agree that [6]

\noindent "{\it But our present [quantum mechanical] formalism is
not purely epistemological; it is a peculiar mixture describing in
part realities of Nature, in part incomplete human information
about Nature | all scrambled up by Heisenberg and Bohr into an
omelette that nobody has seen how to unscramble. Yet we think that
the unscrambling is a prerequisite for any further advance in
basic physical theory. For, if we cannot separate the subjective
and objective aspects of the formalism, we cannot know what we are
talking about; it is just that simple.}"

In this paper we contrast a top-down with an bottom-up
(hidden-variables) description of the illustrious Stern-Gerlach
experiment. Contrasting each other, these descriptions highlight
the quantum structure studies (QSS) as a useful tool in the
foundations and interpretation of quantum theory that is our main
goal. Certain ramifications of the observations made in this paper
will be elaborated elsewhere.

\bigskip

{\bf 2. Outlines of a top-down approach to quantum structures}

\bigskip

This Volume covers the different top-down approaches and
interpretations of quantum structures, see e.g.  [7].
In this section we briefly overview
perhaps the simplest one, which equates "quantum structure" with
the "tensor product structure [of the composite system's Hilbert
space]" [4] (and references therein).

In the universally valid and complete quantum theory\footnote{To be precise, by "universally valid
and complete quantum theory" we have in mind the standard nonrelativistic formalism not extended or amended
by any additional assumptions or interpretational elements.}, every set of
the {\it linearly independent and commuting} degrees of freedom, $\{q_i\}$, defines a
tensor-product structure  for the system's Hilbert space:

\begin{equation}
\mathcal{H} = \otimes_i \mathcal{H}_i,
\end{equation}

\noindent where $\mathcal{H}_i$ concerns the $i$th degree of
freedom\footnote{Notice that a factor-space $\mathcal{H}_i$ may
refer to a particle's spin as a vector observable.}.

By "alternative degrees of freedom" we assume the sets of {\it
arbitrary} degrees of freedom, which are mutually linked via the
invertible linear canonical transformations (LTS), e.g.

\begin{equation}
\xi_m = \sum_i \kappa_{mi} q_i,
\end{equation}

\noindent and analogously for the conjugated momentums (if these
exist). For the unitary matrix $(\kappa_{mi})$ applies the
constraint $\sum_l \kappa_{li} \kappa_{lm} = \delta_{im}$ for real
$\kappa$s.

Then the two sets of the degrees of freedom, $\{q_i\}$ and
$\{\xi_m\}$, define a pair of structures of the composite
system.

To illustrate, consider the paradigmatic example of the hydrogen
atom, which is defined as a pair of quantum particles "electron"
and "proton" via the respective position observables $\vec r_e$,
$\vec r_p$. However, this structure, $e+p \equiv \{\vec r_e, \vec
r_p\}$, is not the only one possible. Rather, it is typically
regarded the alternative atomic "center of mass + relative
particle" ($CM+R$) structure that is defined by the respective
position observables, $\vec R_{CM}$ and $\vec\rho_R$. The
structural transition [8]

\begin{equation}
e+p \to CM+R
\end{equation}

\noindent is due to the (invertible) linear canonical
transformations:

\begin{equation}
\vec R_{CM} = {m_e\vec r_e + m_p\vec r_p \over m_e + m_p}, \vec
\rho_R = \vec r_e - \vec r_p,
\end{equation}

\noindent and with the tensor re-factorization:

\begin{equation}
\mathcal{H}_e \otimes \mathcal{H}_p = \mathcal{H}_{CM} \otimes
\mathcal{H}_R.
\end{equation}

All  kinds and types of the LTS-induced structures are of
interest in the context of the universally valid and complete
quantum theory. For closed quantum systems (subjected to the
Schr\" odinger law), there is no priviledged (preferred)
structure, Section 1. However, for open quantum systems, it is
often conjectured, e.g. [9],  and sometimes justified [8, 10, 11]
existence of a preferred structure (decomposition into subsystems)
due to the environmental influence. Sometimes, the preferred
structure is postulated [12] or expected to exist due to the additional
symmetry-based requirements, see e.g. [13].

In general, {\it locality} is clearly defined for every possible
subsystem of a composite (total) system: Tensor-factorization of
the composite system's Hilbert space uniquely defines {\it  local
observables} for every subsystem. For example, any observable $A$
of the atomic $CM$ system is defined $A_{CM}\otimes I_R$ and, according to eq.(4), is a
"collective" observable regarding the atomic $e+p$ structure.
Equivalently, any observable $B$ for the atomic proton is local
for the $e+p$ structure in the form of $B_p\otimes I_e$ and is a
"collective" observable for the atomic $CM+R$ structure.
Therefore, as long as a measurement of the $A_{CM}\otimes I_R$
observable can induce non-local effects for the $e+p$ structure,
it never--by definition--induces any influence on the atomic $R$
system.

It is essential to note that the tensor-product nonlocality
is more general than nonlocality often regarded in the context of
interpretation of Bell inequalities. An action exerted on e.g. the atomic $CM$ system  {\it only partially
interrupts the atomic electron and proton}. This is a direct consequence of eq.(4):
according to eq.(4), to say that both electron and proton are simultaneously influenced by an action
{\it implies} that  the atomic $CM$ and $R$ systems are also {\it both} simultaneously
influenced by that action. Therefore, a local action on the atomic $CM$ or $R$
systems does not influence either the electron or the proton, except partially, and {\it vice versa}.
Hence we can conclude that the here introduced concept of locality is a prerequisite for the
locality vs. nonlocality studies regarding the Bell inequalities, which assume a fixed
bipartite structure of a composite system.

At this
point, this version of the top-down approach to quantum structures
directly tackles the so-called Tsirelson's problem [14] and also
naturally calls for the  analysis in the context of the so-called
Categorical quantum theory, e.g. [15]--that here will not be
considered. The secondary role of "quantum subsystems", which lack
independent individuality, is exhibited by the so-called "parallel
occurrence of decoherence" on the purely formal level
in the context of the standard environment-induced decoherence theory
[16]. Simultaneous unfolding of the mutually irreducible decoherence
processes regarding different (mutually irreducible) structures of a
single closed system [16] is the price that must be
paid as long as quantum theory is  considered to be universal and complete
in a non-interpretational context.

\bigskip

{\bf 3. The Stern-Gerlach experiment in the universally valid and
complete quantum mechanics}

\bigskip

Now we are prepared to give a top-down description of the
Stern-Gerlach experiment for arbitrary atom not carrying a net
electric charge.

Consider an atom as a set of electrons and the atomic nucleus, which is
typically considered as a point-like (non-structured) quantum
particle. The atom's Hamiltonian:

\begin{equation}
H = \sum_i {\vec p_{ei}^2\over 2m_e} + {\vec p_{nucl}^2\over
2m_{nucl}} - k\sum_i {Ze^2\over\vert \vec x_{ei} - \vec
x_{nucl}\vert} + V
\end{equation}

\noindent with the obvious notation and the $V$ stands for all the
(weaker) Coulomb- and spin-interactions between the particles that
can be treated as the gross perturbation term. Eq. (6) defines the following
tensor-factorization of the atomic Hilbert state-space:

\begin{equation}
\mathcal{H} =
\mathcal{H}_{e}\otimes\mathcal{H}_{n}\otimes\mathcal{H}_{spin},
\end{equation}

\noindent where the atomic spin state-space, $\mathcal{H}_{spin}$, is assumed to be
isomorphic to the single-electron's spin-$1/2$ space and the
remaining factor spaces concern the standard "orbital" (spatial)
degrees of freedom of the electrons ($e$) and of the atomic
nucleus ($n$) systems.

However, the standard theoretical model [5] of the Stern-Gerlach
experiment  regards the alternative "center-of-mass+internal
(relative)" atomic structure for the spatial degrees of freedom.
This structural change\footnote{A generalization of eq.(4).}, $e+n \to CM+R$, induces refactorization of
the Hilbert state-space

\begin{equation}
\mathcal{H} =
\mathcal{H}_{CM}\otimes\mathcal{H}_{R}\otimes\mathcal{H}_{spin}
\end{equation}

\noindent as well as the alternative form of the atomic
Hamiltonian, $H = {\vec P_{CM}^2\over 2M} + \sum_i \vec
p_{Ri}^2/2\mu_i + V_{Coul} + V$ where the $\mu_i$s represent the
"reduced masses" while $V_{Coul}$ is the nucleus-induced classical
Coulomb field for the $R$ system's degrees of freedom [4, 8].
Neglecting the weak term $V$ and bearing in mind eq.(3), the
Hamiltonian exhibits the variables separation:

\begin{equation}
H = {\vec P_{CM}^2\over 2M}\otimes I_R\otimes I_{spin}  + I_{CM}
\otimes H_R \otimes I_{spin},
\end{equation}

\noindent where the internal atomic energy $H_R = \sum_i \vec
p_{Ri}^2/2\mu_i + V_{Coul}$.

Placing the atom in a sufficiently strong magnetic field along the
$z$-axis is modelled by the interaction term for the $CM+spin$
system [5]:

\begin{equation}
H_{CM+spin} = - \vec\mu_{spin}\cdot \vec B(Z_{CM}),
\end{equation}

\noindent where $\vec \mu_{spin} = - \mu \vec S$ is the
atomic-spin magnetic dipole; if we apply the standard
approximation $\vec \mu_{spin} \approx \vec \mu_e$, then the
constant $\mu = \mu_B$ is the standard "Bohr magneton" and $\vec
S$ is the spin of the electron.

From eqs.(9), (10) it directly follows that the unitary dynamics
of the atom induces quantum entanglement between the atomic $CM$
and "spin" systems, while the absence of interaction between the
$R$ system with the $CM$ and the "spin" system allows for the
"separation" of the $R$ system's state from the rest in the
form of:

\begin{equation}
\vert\Psi\rangle = \left(\sum_i c_i \vert
i\rangle_{CM}\otimes\vert i\rangle_{spin}\right) \otimes \vert
\chi\rangle_R,
\end{equation}

\noindent in an instant of time $t$; for the spatial $i = +, -$,
[indicating the above and below the $XY$-plane],  the spin
states $i = \downarrow, \uparrow$, respectively, and
$c_i=2^{-1/2}$.

Bearing in mind Section 2, the physical meaning of eq.(11) is
rather obvious:

\noindent ($\mathcal{S}$) {\it Magnetic field} locally {\it acts
on the atomic $CM$ system, while leaving the atomic internal (the
$R$) system intact.}

\noindent To the extent that eq.(11) captures the phenomenology,
the same can be told for the statement $\mathcal{S}$.

This trivial observation carries a non-trivial content regarding
the atomic $e+n$ structure. Actually, bearing in mind Section 2,
the $\mathcal{S}$ statement straightforwardly implies the
following observation:

\noindent ($\mathcal{O}$) {\it Neither the electrons} ($e$) {\it
nor the atomic nucleus} ($n$) {\it are influenced by the magnetic
field}.

The statement $\mathcal{O}$ is easily proved as a repetition of the arguments of Section 2: Assume that the
magnetic field "sees" both the atomic electrons and the atomic
nucleus. Then  an influence exerted on the electrons positions,
$\vec r_i$, {\it and} on the nucleus position, $\vec r_n$, in analogy with
eq.(4), directly gives rise to the conclusion that {\it both} the
atomic $CM$ and $R$ systems are influenced by the magnetic
field--in contradiction with eq.(11)  i.e. with  $\mathcal{S}$.

Therefore, equal status of every (physically reasonable) structure
of a composite system {\it implies} the purely local influence of
the magnetic field. Furthermore, this influence cannot be
precisely described for the atomic $e+n$ structure as, due to
$\mathcal{S}$ and in analogy with eq.(4), both $e$ and $n$ are
only "partially" seen by the magnetic field.

\bigskip

{\bf 4. A hidden-variables description of the Stern-Gerlach
experiment}

\bigskip

Influence of the magnetic field on the atomic $e$ and $n$ systems
cannot be imagined in the universally valid and complete quantum
theory--the statement $\mathcal{O}$ of Section  3. Hence assuming

\noindent
($\mathcal{S}'$) {\it Magnetic field locally acts on the
atomic} $e$ {\it  and} $p$ {\it systems, independently.}

\noindent
directly leads to the conclusion that there are certain
{\it hidden variables} in the description of the atom in the
Stern-Gerlach experiment. That is, a "non-hidden" influence of the magnetic field
implies the above statement $\mathcal{S}$, i.e. the $\mathcal{O}$ of the  previous section, not to be correct.

It is natural to assume that the electrons-system and the nucleus
positions, $\vec r_i$ and $\vec r_n$, respectively, can play the
roles of "hidden variables" (HV). Assuming the ontic status of
those variables like in the Bohm's theory [5], it is possible
qualitatively to devise a scenario in which the standard
description, eq.(11), applies and still to describe the influence
of the field on {\it both} the $e$ and $n$ atomic systems. For
simplicity, consider the hydrogen atom described by eq.(11). Then
it's easy to imagine that the external magnetic field $\vec B$
drives both the electron and the proton so as to have the
dynamical change of the $CM+spin$ system as described by eq.(11),
while the relative distance between the $e$ and $p$ is determined
by the probability density $\vert\chi(\vec\rho_R)\vert^2$; $\vec
\rho_R = \vec r_e - \vec r_p$, cf. Section 2 for the notation.

Needless to say, in this picture, the alternative atomic structure
$CM+R$ is  artificial--simply a mathematical artifact without any
physical meaning. Just like in classical physics, this structure
can be used to ease mathematical manipulation, while all the
physically relevant results must be  expressed in the terms of the
fundamental degrees of freedom, of $\vec r_e$ and $\vec r_p$.

The magnetic field eq.(10), which is of the form $\vec
B(Z_{CM}\otimes I_{spin})$ for the $CM+R$ structure, can represent an
interaction term for the atomic $e+p$ structure: $\vec B\left((m_e
z_e\otimes I_p + m_p I_e\otimes z_p)/M\right)$; $M=m_e+m_p$. This
interaction would produce entanglement for the $e$ and $p$ even if
there were not the Coulomb interaction in the $e+p$ structure.
That is, regarding the atomic $e+p$ structure, eq.(11) takes the
form\footnote{The adiabatic cut of the electron from the proton
due to $m_e/M\ll 1$ would not change our argument even if
numerically justified, see Ref. [4] for more details.}:

\begin{equation}
 \sum_i c_i \vert \psi_i\rangle_{e+p} \otimes
\vert \chi_i\rangle_{spin} = \vert \Psi\rangle = \sum_i c_i \left( \sum_{\alpha}
d_{\alpha}^i \vert \alpha\rangle_e \otimes \vert \alpha\rangle_p
\right) \otimes \vert \chi_i\rangle_{spin},
\end{equation}

\noindent as an example of Entanglement Relativity [4, 18] (and references therein).

By definition, eq.(12) is {\it insensitive} to the fundamental ({\it subquantum})
influence of the magnetic field on the atomic $e$ and $p$
subsystems and can be assumed to give rise to the subquantum (HV)
probability density with purely classical correlations
[that are induced by the Coulomb and/or the external magnetic field], of the
general form of:

\begin{equation}
\mu(\lambda) = \sum_i p_i f_{1i}(\vec r_e) f_{2i}(\vec r_p).
\end{equation}

\noindent If normalized, the local probability densities $f_i$s
imply $\sum_i p_i = 1$, while $\int \mu(\lambda) d\lambda = 1$.

Hence a HV theory \' a la Bohm can be summarized by the following
observation while taking over the meaning of "locality" from
Section 3:

\noindent ($\mathcal{O}'$) {\it Both the electrons} (e) {\it and the atomic
nucleus} (n) {\it systems are} locally {\it  influenced by the
magnetic field so as that cannot be observed on the quantum level
of} eq.(12).

\bigskip

{\bf 5. Discussion}

\bigskip

Quantum structural studies (QSS) of Section 2 are {\it minimalist} in that they neither extend nor interpret
the standard quantum mechanical formalism. In this sense, QSS introduce a non-standard methodology in the standard
quantum theory with the following benefits.

First, the very basic QSS concept of {\it locality}, Section 2, is
a prerequisite of the locality (i.e. of "local causality") \' a la Bell [17]
and is compatible with the "non-locality"
in a hidden-variables theories \' a la Bohm [17]--cf. ($\mathcal{O}'$), Section 4. Therefore different
contents of "locality" are naturally linked through the quantum structural studies.

Second, on the {\it ontological level}, QSS as per Section 2 is in sharp contrast with the reductionistic interpretations of quantum theory
\' a la Bohm [5, 17]. Distinguishing between these two approaches to quantum theory may be seen as an amendment to the
tests of the Bell inequalities [17]: one should simultaneously consider more than one composite-system's structure,
which leads to the following observation.

Third, all the structural considerations are by definition {\it contextual}. Regarding the atom as a whole,
all  observables of every single atomic subsystem are the atomic observables too. However,
measurements that are local relative to one atomic structure need not imply any information
regarding an alternative atomic structure. Capturing the atoms impinging on a screen directly reveals the
     atomic $CM$ position on the screen but does not provide any
     information regarding the atomic $R$ system. Thus
     the Stern-Gerlach experiment is local relative to the $CM+R$
     structure and does not provide much information regarding the atomic
     $e+p$ structure. Therefore, the
     concept of contextuality is also extended, i.e. in a  sense
     generalized: it does not assume exclusively a single (fixed) structure of
     a composite system and also allows for the mutually compatible observables, cf. e.g. eq.(4).

Fourth, while eqs.(12) and (13) are  compatible, they are still limited from the point of view of the quantum structural studies.
Simultaneous consideration of the physically fundamental and artificial structures pose specific constraints on the
HV theories as per Section 3. To see this, consider the asymptotic limit of the hydrogen atom--i.e. a pair of the free noninteracting electron and proton--out of any external field.
The "natural" choice of the classical density probability

\begin{equation}
\mu(\lambda) = f_1(\vec r_e) f_2(\vec r_p)
\end{equation}

\noindent implies a conflict with the standard task of introducing the artificial $CM+R$ structure, eq.(4).
Actually, as a classical counterpart of the quantum correlations relativity [4,  18], typically:

\begin{equation}
 f_1(\vec r_e) f_2(\vec r_p) = \mu(\lambda) = \sum_i q_i g_{1i}(\vec R_{CM}) g_{2i}(\vec \rho_R) \neq G_1(\vec R_{CM}) G_2 (\vec \rho_R),
\end{equation}

\noindent which is in contrast with the standard task of "variables separation" (and integrability) in classical mechanics.
That is, mutually separated $CM$ and $R$ systems are typically described by the rhs of eq.(15).

Conversely,  the inverse of eq.(14):

\begin{equation}
 G_1(\vec R_{CM}) G_2 (\vec \rho_R) = \mu'(\lambda) = \sum_i p_i f_{1i}(\vec r_e) f_{2i}(\vec r_p) \neq f_1(\vec r_e) f_2(\vec r_p),
\end{equation}

\noindent challenges the above-distinguished "natural" choice of eq.(14).

Thereby a choice for the  probability density $\mu(\lambda)$ becomes not as free as it may seem in the standard HV theories. Despite the fact that the alternative structures are artificial,
the simple {\it mathematical} considerations as per eqs. (13)-(16) confine the considerations in a non-trivial way yet fully to be explored.
In the context of the Stern-Gerlach experiment, this choice should be accompanied with eq.(11) and eq.(12) while bearing in mind that eq.(11) describes the
{\it operationally accessible} [2] situation.

It is worth repeating: Structure of a composite system is conditional in the universally valid and complete quantum theory, Section 2. As distinct from the
prevailing reductionistic wisdom, the quantum structural studies of Section 2 highlight existence
of a preferred structure of an {\it open} quantum system as a {\it relational} concept.
There is nothing in the formalism as well as in the physical ontological basis that could
suggest the ontologically superior status of any structure of a closed composite quantum system.
Thus the quantum structural studies as per Section 2 directly challenge the standard physical
conceptualization embodied in the concept of "elementary particles" [4, 17]:
{\it In the non-relativistic domain, quantum formalism can properly work even without assuming
the ontological status of elementary particles.}

This brings us to the following mathematical forms of the basic task of the quantum structural studies.
On one hand, it's the so-called Tsirelson's problem [14]: Whether or not a composite system's
state space is built via the tensor-product structures as assumed in QSS? And the corollary: What if the tensor-product structures have an alternative--what
is the meaning of "locality" in that new, hypothetical discourse? As another natural frame for the quantum structural studies we
emphasize the Category studies [15], in which {\it correlations} are primary and "{\it correlata}" (subsystems)
are secondary. This is a natural framework for the emerging approach of "there are no particles" to quantum
foundations [4, 19].

From the purely operational point of view, QSS elevate the following seemingly "philosophical" issue:
How can we be sure in a concrete physical situation which degrees of freedom of a composite system have been
targeted by our apparatus(es)?
The natural assumption that acquiring information about a composite system is limited to a
"small" set of the system's degrees of freedom {\it implies} locality as per Section 2.
On one hand, measurement of the proton's position in the hydrogen atom gives a close value for the atomic $CM$ position but introduces entanglement
for the $CM$ and $R$ systems, cf. eq.(4), that in principle can be experimentally tested.
Hence non-equivalence of measurements of the proton and the $CM$ system's positions that instantiates locality of Section 2.
On the other hand, targeting {\it both} the $e$ and $p$ atomic systems      also
provides an  information regarding the positions of the formal systems $ \vec R_{CM}$  and $\vec \rho_R$
[contextual measurements with only linear increase in uncertainties for $\vec R_{CM}$  and $\vec \rho_R$].
Hence the "total" measurements--of the pair $ \vec r_{e}, \vec r_p$ (or of the pair $ \vec R_{CM}, \vec \rho_R$)--on the atom are practically equivalent.

Now the point strongly to be stressed is that quantum uncertainty, i.e. the non-increase of information regarding the related conjugate momentums,
may appear as a kind of {\it limitation} in the physical process of {\it measurement} (that is understood as the process of acquiring information),
not necessarily of the fundamental physical description of the atomic
$e$ and $p$ subsystems. In effect, the quantum part
of the theory, eq.(12), may be {\it not a fundamental} but an emergent description of the quantum system called "atom" {\it due
to the process of acquiring information} now crying for explanation and called "measurement".
In our opinion, the dichotomy of "particles + quantum field" [5] is a suitable phrase, not a satisfying explanation yet.
This position is in contrast to that of Bell's [20], which discards foundational importance of the quantum measurement process.
To this end, setting "measurement" as a purely operational
concept [21, 22] makes the two competing theories (Bohm [5] versus Qubism [22]) virtually indistinguishable
everywhere except possibly on the operationally-inaccessible ontological level.

Operationally again, QSS are sometimes reduced to the topic of experimental accessibility of the composite system's observables [2]. While
"accessibility" assumes the preferred tensor-product structure, which may be environment-induced [8-11], there appear the following subtleties.
On one hand, not all  observables of a subsystem pertaining to a preferred structure of the composite system are on the equal physical footing.
To this end, the symmetry of the open system's effective Hamiltonian [23] may be decisive--not only a choice between the noncommuting e.g. "position" and "momentum" (or energy)
observables should be made, but also a choice between the commuting observables defining "object" (e.g. the Descartes vs spherical degrees of freedom) are
of interest [24]. On the other hand, the environmental influence is typically approximate and hence one may expect some "emergent" degrees of freedom
[25].

This brings us to the following answers to the topical questions of this Volume. The familiar classical
center-of-mass degrees of freedom are environment-induced, relationally realistic and operationally accessible
degrees of freedom, which do not require any ontologically superior subquantum degrees of freedom. Hence the transition from quantum to classical regards the specific,
{\it local} degrees of freedom that, we believe, may be at the core of the other aspects of the problem of "transition from quantum to classical" [9]. In other
words, a part of the problem may reside on the assumption that the problem must be solved in the terms of the "fundamental"
(ontological) structure of a composite system. Unfortunately, we are not aware nor do we offer an elaborated experimental proposal in this regard.

While quantum structural studies challenge [26] the interpretations \' a la Everett [25, 27], the discourses  regarding the
non-universally valid quantum theory [28] as well as  "the problem of time" [29-31] (and references therein) are in order.
To this end, the work is in progress and the results will be presented elsewhere.

\bigskip

           \bigskip

           {\bf Acknowledgements} We benefited much from
           discussions with Ruth Kastner. Financial support from
           Ministry of Science Serbia grant no 171028 is
           acknowledged, as well as for MD by the ICTP - SEENET-MTP project PRJ-09 Cosmology and Strings in frame of the SEENET-MTP Network.

\bigskip

[1] S. Y. Auyang, {\it Foundations of Complex-system Theories}, Cambridge University Press, Cambridge, 1998

[2] P. Zanardi, Virtual quantum subsystems. {\it Phys. Rev. Lett.} {\bf 87}, 077901 (2001)

[3] J. Halliwell, Emergence of hydrodynamic behaviour. In
 {\it Many
Worlds? Everett, Quantum Theory, and Reality}, S. Saunders, J. Barrett, A. Kent, D. Wallace (eds.),  Oxford
University Press, Oxford, 2010, pp. 99-117.

[4] J. Jekni\' c-Dugi\' c, M. Arsenijevi\' c, M. Dugi\' c,
    {\it Quantum Structures. A View of the Quantum World}, LAP Lambert
    Scientific Publishing, Saarbr\" ucken, 2013; also available at
    e-print arXiv     arXiv:1306.5471 [quant-ph]

[5] D. Bohm, {\it Quantum Theory}, Prentice Hall, New York, 1951

[6] E. T. Jaynes, In {\it Complexity, Entropy, and the Physics of
Information}, W. H. Zurek (ed.), Addison-Wesley, 1990, p. 381

[7]    S. Fortin, O. Lombardi, A top-down view of the classical limit of quantum mechanics,
forthcoming in Kastner R E, Jekni\' c-Dugi\' c J, Jaroszkiewicz G, eds., {\it Quantum Structural
Studies}. Singapore: World Scientific Publishers, 2016.

[8] J. Jekni\' c-Dugi\' c, M. Dugi\' c, A. Francom, M. Arsenijevi\' c,
Quantum Structures of the Hydrogen Atom, {\it Open Access Library Journal} {\bf 1}, e501
(2014)

[9] D. Giulini, E. Joos, C. Kiefer, J. Kupsch, I.-O. Stamatescu, H. D. Zeh,
{\it Decoherence and the appearance of a classical world
    in quantum theory}, Spinger, Berlin, 1996

[10] M. Arsenijevi\' c, J. Jekni\' c-Dugi\' c, M. Dugi\' c,
Asymptotic dynamics of the alternate degrees of freedom for a
two-mode system: an analytically solvable model, {\it Chinese Phys. B}
{\bf 22}, 020302 (2013)

[11] A. Stokes, A. Kurcz, T. P. Spiller, A. Beige, Extending the
validity range of quantum optical master
 equations, {\it Phys. Rev. A} {\bf 85}, 053805  (2012)

[12] W. H. Zurek, Decoherence, einselection, and the quantum origins of the classical,
{\it Rev. Mod. Phys.} {\bf 75}, 715 (2003)

[13]     N. L. Harshman, Symmetry and Natural Quantum Structures for Three-Particles
                         in One-Dimension, forthcoming in Kastner R E, Jekni\' c-Dugi\' c J, Jaroszkiewicz G, eds., {\it Quantum Structural
Studies}. Singapore: World Scientific Publishers, 2016.

[14] B. S. Tsirelson, Some results and problems on quantum Bell-type
     inequalities, {\it Hadronic J. Suppl.} {\bf 8}, 329–345 (1993)

[15] B. Coecke, A. Kissinger,  Categorical Quantum Mechanics
I: Causal Quantum Processes, e-print arXiv     arXiv:1510.05468
[quant-ph]

[16] M. Dugi\' c, J. Jekni\' c-Dugi\' c,  Parallel decoherence in
composite quantum
 systems, {\it Pramana} {\bf 79}, 199       (2012)

[17] H. M. Wiseman, {\it J. Phys. A: Math. Theor.} {\bf 47}, 424001 (2014)

[18] M. Dugi\' c, M. Arsenijevi\' c, J. Jekni\' c-Dugi\' c,
Quantum correlations relativity, {\it Sci. China, Phys. Mech. Astron.}
{\bf 56}, 732                          (2014)

[19] H. D. Zeh, There are no Quantum Jumps, nor are there Particles, {\it Phys. Lett. A} {\bf 172},
189 (1993)

[20] J. S. Bell, Against "Measurement", In {\it Sixty-Two Years of Uncertainty
                       Historical, Philosophical, and Physical Inquiries into the Foundations of Quantum
                       Mechanics}, Arthur I. Miller (ed),  NATO ASI Series
                                             Volume 226 1990, pp. 17-31

[21] M. A. Nielsen, I. L. Chuang, {\it Quantum Computation and Quantum Information},
     Cambridge University Press, Cambridge, 2000

[22] C. A. Fuchs, N. David Mermin, R. Schack, An Introduction to QBism
     with an Application to the Locality of Quantum Mechanics, {\it Am. J.
     Phys.}
     {\bf 82}, 749 (2014)

[23]  N. L. Harshman, S. Wickramasekara,  Galilean and Dynamical Invariance of Entanglement in Particle Scattering
{\it Phys. Rev. Lett.} {\bf 98}, 080406 (2007)

[24] W. H. Zurek, Preferred States, Predictability, Classicality and the Environment-Induced Decoherence,
{\it Prog. Theor. Phys.} {\bf 89}, 281 (1993)

[25] D. Wallace, {\it Emergent Multiverse: Quantum Theory According to the
     Everett Interpretation}, Oxford University Press, Oxford, 2012

    [26] J. Jekni\' c-Dugi\' c, M. Dugi\' c, A. Francom, Quantum
     Structures of a Model-Universe: Questioning the Everett
     Interpretation of Quantum Mechanics, {\it Int. J. Theor. Phys.} {\bf
     53}, 169

[27] S. Saunders, J. Barrett, A Kent, D. Wallace (Eds.), {\it Many Worlds?
     Everett, Quantum Theory, and Reality}, Oxford University Press, Oxford,
     2010.

[28] G. C. Ghirardi, A. Rimini, T. Weber, Unified dynamics for microscopic and macroscopic systems,
{\it Phys. Rev.
D} {\bf 34}, 470, 1986

[29] J. Jekni\' c-Dugi\' c, M. Arsenijevi\' c, M. Dugi\' c, A local-time-induced pointer
     basis, {\it Proc. R. Soc. A} {\bf 470}, 20140283 (2014) .

[30] J. Jekni\' c-Dugi\' c, M. Arsenijevi\' c, M. Dugi\' c,  Dynamical emergence of Markovianity in Local Time Scheme,
e-print arXiv arXiv:1510.02913 [quant-ph]

[31] C. Baumgarten, Minkowski Spacetime and QED from Ontology of Time,
forthcoming in Kastner R E, Jekni\' c-Dugi\' c J, Jaroszkiewicz G, eds., {\it Quantum Structural
Studies}. Singapore: World Scientific Publishers, 2016.

\end{document}